\documentclass[pra, twocolumn, twoside, a4paper, fleqn, floatfix]{revtex4}
\usepackage{times}
\usepackage{graphicx,units}
\usepackage{amssymb,stmaryrd}
\usepackage{pifont,wasysym}
\newcommand{\Diamondblack}{\ding{117}}
\newcommand{\medbullet}{\ding{108}}
\newcommand{\medcirc}{\ocircle}
\usepackage{natbib}
\usepackage[usenames,dvipsnames]{color}

\usepackage[bookmarks=true,letterpaper=true,colorlinks=true,urlcolor=blue,linkcolor=blue,citecolor=blue]{hyperref} 
\received{January 8, 2009}
\published{June 19, 2009}
\begin{document}

\title{Stability and Electronic Properties of TiO${}_{\text{2}}$
  Nanostructures With and Without B and N Doping}   
\author{D. J. Mowbray}
\email{dmowbray@fysik.dtu.dk}
\author{J. I. Martinez}
\author{J. M. Garc{\'{\i}}a-Lastra}
\author{K. S. Thygesen}
\author{K. W. Jacobsen}
\affiliation
{Center for Atomic-scale Materials Design, Department of Physics,
  Technical University of Denmark, DK-2800 Kgs.~Lyngby, Denmark} 

\begin{abstract}
We address one of the main challenges to TiO${}_{\text{2}}$-photocatalysis,
namely band gap narrowing, by combining nanostructural changes with
doping. With this aim we compare TiO${}_{\text{2}}$'s electronic properties for
small 0D clusters, 1D nanorods and nanotubes, 2D layers, and 3D
surface and bulk phases using different approximations within density
functional theory and GW calculations. In particular, we propose very
small ($R\lesssim$ 5 \AA) but surprisingly stable nanotubes with
promising properties. The nanotubes are initially formed from
TiO${}_{\text{2}}$ layers with the PtO${}_{\text{2}}$ structure, with the smallest (2,2)
nanotube relaxing to a rutile nanorod structure.  We find that quantum
confinement effects -- as expected -- generally lead to a widening of
the energy gap. However, substitutional doping with boron or nitrogen
is found to give rise to (meta-)stable structures and the introduction
of dopant and mid-gap states which effectively reduce the band gap.
Boron is seen to always give rise to $n$-type doping while depending
on the local bonding geometry, nitrogen may give  rise to $n$-type or
$p$-type doping. For undercoordinated TiO${}_{\text{2}}$  surface structures
found in clusters, nanorods, nanotubes, layers and surfaces nitrogen
gives rise to acceptor states while for larger clusters and bulk
structures donor states are introduced.     
\end{abstract}
\maketitle

\section{Introduction}

Motivated by the world's ever increasing need for cleaner burning
fuels and more viable forms of renewable energy, hydrogen production
via photocatalysis has been intensely researched as a possible
candidate for addressing these issues. Since the first experimental
formation of hydrogen by photocatalysis in the early
1980s~\cite{FirstPhotocatalysis}, TiO${}_{\text{2}}$ has been the catalyst of
choice.  Reasons for this include the position of TiO${}_{\text{2}}$'s
conduction band above the energy of hydrogen formation, the relatively
long lifetime of excited electrons which allows them to reach the
surface from the bulk, TiO${}_{\text{2}}$'s high corrosion resistance compared
to other metal oxides, and its relatively low cost
\cite{Gratzel2001,Hoffmann1995,Khan2002}.    

However, the large band gap of bulk TiO${}_{\text{2}}$ ($\approx$ 3 eV)
means that only high energy UV light may excite its electrons.  This
effectively blocks most of the photons which pierce the atmosphere,
typically in the visible range, from participating in any bulk
TiO${}_{\text{2}}$ based photocatalytic reaction.  On the other hand,
the difference in energy between excited electrons and holes, i.e. the
band gap, must be large enough ($\gtrsim$ 1.23 eV) to dissociate water
into hydrogen and oxygen.  For these reasons it is of great interest
to adjust the band gap $\varepsilon_{gap}$ of TiO${}_{\text{2}}$ into the range
1.23 $\lesssim \varepsilon_{gap} \lesssim$ 2.5 eV, while maintaining
the useful properties mentioned above \cite{TiO2PRL}.  

With this aim, much research has been done on the influence of
TiO${}_{\text{2}}$ nanostructure~\cite{Structures,ExpTiO2clusters,TiO2NTs1,
  TiO2NTs2,Delaminated} and dopants
\cite{JACS-suil,N-TiO2NTarraydoping,N-TiO2doping,N-TiO2NTdoping,
N-TiO2NTdoping2,Nambu,Graciani,TiO2PRL}     
on photocatalytic activity.  For low dimensional nanostructured
materials, electrons and holes have to travel shorter distances to
reach the surface, allowing for a shorter quasi-particle
lifetime. However, due to quantum confinement effects, lower
dimensional TiO${}_{\text{2}}$ nanostructures tend to have  
\emph{larger} band gaps \cite{TiO2Rev1}.  On the other hand, 
although doping may introduce mid-gap states, recent
experimental studies have shown that boron and nitrogen doping of bulk
TiO${}_{\text{2}}$ yields band gaps \emph{smaller} than the threshold
for water splitting \cite{JACS-suil,N-TiO2NTarraydoping}.  
This suggests that low dimensional structures with band gaps larger
than about 3.0 eV may be a better starting point for doping. 

The experimental synthesis and characterization of nanostructured
materials is in general a costly and difficult task.  On the other
hand, modern electronic structure modelling has reached a level where
large-scale calculations can provide realistic descriptions of
structure and electronic properties. 

Based on our investigation, we suggest as a promising candidate small
($R \lesssim$ 5 \AA) TiO${}_{\text{2}}$ nanotubes, with a hexagonal ABC PtO${}_{\text{2}}$
structure (HexABC), which we find to be surprisingly stable, even in
the boron and nitrogen doped forms.   This stability may be attributed to
  their structural similarity to bulk rutile TiO${}_{\text{2}}$, with the
  smallest nanotube having the same structure as a rutile nanorod.

A further difficulty for any  photocatalytic system is
controlling how electrons and holes travel through the system
\cite{Turner}. For this reason, methods 
for reliably producing both $n$-type and $p$-type TiO${}_{\text{2}}$
semiconducting materials are highly desirable. So far, doped TiO${}_{\text{2}}$
tends to yield only $n$-type semiconductors.  We propose that $p$-type
TiO$_{\text{2}}$ semiconducting materials may be obtained by nitrogen doping
surface sites of low dimensional materials.

In this study we report the results of density functional theory (DFT)
and GW calculations of the energetic stability and electronic structure 
of recently suggested (TiO${}_{\text{2}}$)${}_{n}$ clusters ($n\leq$ 9)
\cite{Structures} and novel ($n$,$n$) TiO${}_{\text{2}}$ nanotubes ($n\leq$ 4)
in the undoped, boron doped, and nitrogen doped forms. The formation
energy $E_{f\!orm}$, density of states (DOS), and energy gap
$\varepsilon_{gap}$, for these systems are compared with that for 2D
HexABC and anatase layers, and 3D TiO${}_{\text{2}}$ anatase surface, anatase
bulk, and rutile bulk phases. Furthermore, we analyze how boron and
nitrogen doping influences the DOS for these systems, and how their
nanostructure may determine whether the resulting semiconductor is
$n$-type or $p$-type.  

In Sec.~\ref{methodology} we describe the DFT and GW methodologies
used to obtain the energies and electronic structure of the studied
systems.  We compare the energetic stability of the systems in
Sec.~\ref{stability}. In Sec.~\ref{electronic} we discuss the
electronic structures of the systems, showing the DOS and energy gaps,
followed by a concluding section.  

\section{Methodology}\label{methodology}

All DFT calculations have employed the RPBE exchange correlation
(xc)-functional~\cite{RPBE}. The plane-wave code
\textsc{dacapo}\cite{Dacapo,ASE} was used for structural minimization,
the real-space code \textsc{octopus}\cite{Octopus} for charged
calculations, and \textsc{yambo}\cite{Yambo} with
\textsc{abinit}\cite{Abinit} or \textsc{PWscf}\cite{pwscf} for GW
calculations. A plane-wave cutoff of 340 eV was used, with a
Monkhorst-Pack {\textbf{k}}-point sampling of  1$\times$1$\times$12
for TiO${}_{\text{2}}$ nanotubes, where the nanotube axis is parallel to the
$z$-axis, 12$\times$12$\times$1 for TiO${}_{\text{2}}$ layers and surfaces,
where the normal direction is parallel to the $z$-axis, and
12$\times$12$\times$12 for TiO${}_{\text{2}}$ bulk phases. All structures have
been relaxed until a maximum force below 0.04 eV/\AA\ was obtained.
The occupation of the one electron states was calculated at a
temperature of $k_B T \approx $0.1 eV for the periodic systems and
$k_B T \approx $0.01 eV for the clusters, with all energies
extrapolated to $T = $0 K. Spin unpolarized calculations have been
performed for the undoped TiO${}_{\text{2}}$ systems, while spin polarized
calculations have been performed for all doped TiO${}_{\text{2}}$ systems,
since the unit cells for the doped systems contain an odd number of
electrons.

Doping of (TiO${}_{\text{2}}$)${}_n$ clusters has been modeled by
substituting a single boron or nitrogen atom in each geometrically
inequivalent oxygen site of the most stable isomer to determine the
most stable doping site.  Only clusters of sufficient
size to obtain experimentally realizable doping fractions
\(\lesssim{10}\)\% (${5}\leq n \leq{9}$) have been considered
\cite{N-TiO2doping,N-TiO2NTdoping,N-TiO2NTarraydoping,N-TiO2NTdoping2,
JACS-suil}.      

To model TiO${}_{\text{2}}$ nanotube doping, we have repeated
the minimal unit cell four times along the tube axis, and substituted
a single boron or nitrogen atom in each geometrically inequivalent
oxygen site to obtain the most stable doped structure. In this way,
dopant-dopant interactions are minimized by maintaining a dopant
separation of approximately 12 \AA. This corresponds to experimentally
realizable doping fractions of 3.1\%, 2.1\%, and 1.6\% for TiO${}_{\text{2}}$
(2,2) nanorods, (3,3) nanotubes and (4,4) nanotubes, respectively.   

Doping of TiO${}_{\text{2}}$ layers, surfaces, and bulks has been similarly
modeled by repeating the minimal unit cell twice in each periodic
direction, and substituting a single boron or nitrogen atom in each
geometrically inequivalent oxygen site to obtain the most stable
structure.  Experimentally realizable doping fractions of 5.6\%,
3.1\%, and 3.1\% were thus obtained for the layers, surfaces, and
bulks respectively. 

\section{Energetic Stability}\label{stability}

\begin{figure}
\includegraphics[width=\columnwidth]{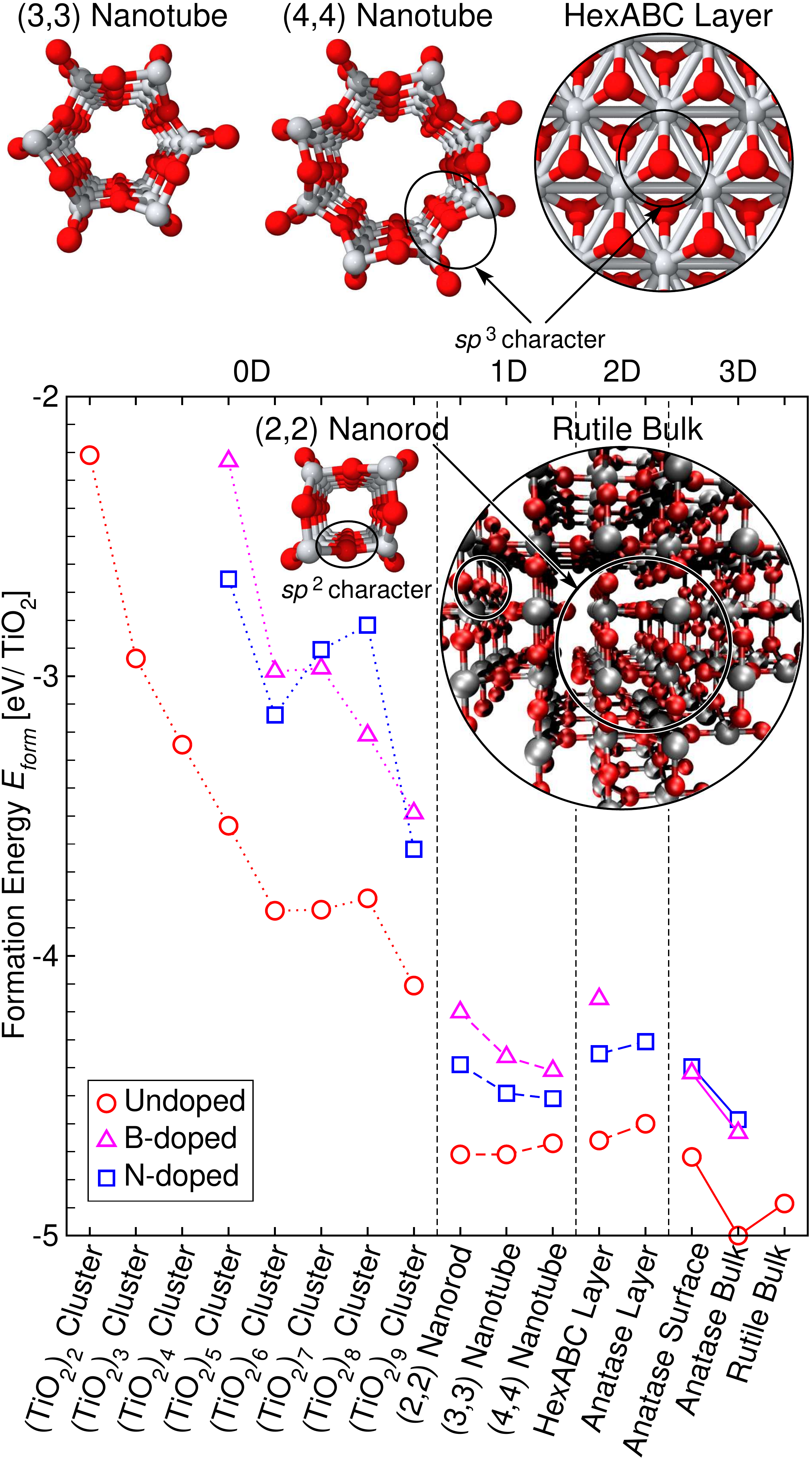}
\caption{
Formation energy $E_{f\!orm}$ in eV per TiO${}_{\text{2}}$ functional unit versus
TiO${}_{\text{2}}$ structure for 0D (TiO${}_{\text{2}}$)${}_n$ clusters ($n\leq \text{9}$),
1D TiO${}_{\text{2}}$ (2,2) nanorods, (3,3) nanotubes, and (4,4) nanotubes,
2D HexABC and anatase layers, and 3D anatase surface, anatase bulk,
and rutile bulk phases. DFT calculations using RPBE for the undoped
({\color{red}{$\medcirc$}}, red), nitrogen doped
({\color{blue}{$\boxempty$}}, blue), and boron doped
({\color{magenta}{$\vartriangle$}}, magenta) systems are shown. Note
the $sp^{\text{2}}$ character of oxygen in the more stable rutile and (2,2)
nanorod structure, and the $sp^{\text{3}}$ character of oxygen in the less
stable layer and larger nanotube structures.}\label{Fig1} 
\end{figure}

We define the formation energy $E_{f\!orm}$ for a given structure
consisting of $n$ TiO${}_{\text{2}}$ functional units as 
\begin{eqnarray}
E_{f\!orm}&=&
\frac{1}{n}E[\text{Ti}_n\text{O}_{\text{2}n-p-q}\text{B}_p\text{N}_q]
- E[\text{TiO}_{\text{2}}]\nonumber\\ 
&& -\frac{1}{n}\left(pE[\text{B}] + qE[\text{N}] -
  (p+q)E[\text{O}]\right) ,
\end{eqnarray}
where $E[\text{Ti}_n\text{O}_{\text{2}n-p-q}\text{B}_p\text{N}_q]$ is
the total energy for the system, $E[\text{TiO}_{\text{2}}]$ is the
energy of an isolated TiO${}_{\text{2}}$ molecule, and $E[\text{O}]$,
$E[\text{B}]$, and $E[\text{N}]$ are the respective energies with
reference to gas phase species for O, B, and N.  These have been
obtained using the experimental doping reactions
\begin{eqnarray}
E[\textrm{O}] &=& E[\textrm{H}{}_{\text{2}}\textrm{O}]
-E[\textrm{H}{}_{\text{2}}] -\Delta
H[\textrm{H}_{\text{2}}\textrm{O}],\\  
E[\textrm{B}] &=&
\frac{{1}}{\text{2}}\left(E[\textrm{B}_{\text{2}}\textrm{H}{}_{{6}}]-
  3E[\textrm{H}_{\text{2}}]-\Delta
  H[\textrm{B}_{\text{2}}\textrm{H}_{\text{3}}]\right),\\  
E[\textrm{N}] &=& \frac{{1}}{\text{2}}E[\textrm{N}_{\text{2}}], 
\end{eqnarray} 
where the formation reaction enthalpies are $\Delta
H$[H${}_{\text{2}}$O] $\approx -$2.506 eV and $\Delta H$[B${}_{\text{2}}$H${}_{\text{6}}$]
$\approx$ 0.377 eV, as taken from Ref.~\cite{CRCHandbook}. This avoids
difficulties associated with modeling isolated atoms and the triplet
state of molecular oxygen.   

Fig.~\ref{Fig1} shows DFT calculated formation energies
$E_{f\!orm}$ for undoped, boron doped, and nitrogen doped forms of 0D (TiO${}_{\text{2}}$)${}_n$
clusters ($n\leq \text{9}$), 1D ($n$,$n$) nanotubes ($n \leq
{4}$), 2D HexABC and anatase layers, and
3D anatase surface, anatase bulk, and rutile bulk phases.

\begin{figure}
\includegraphics[width=\columnwidth]{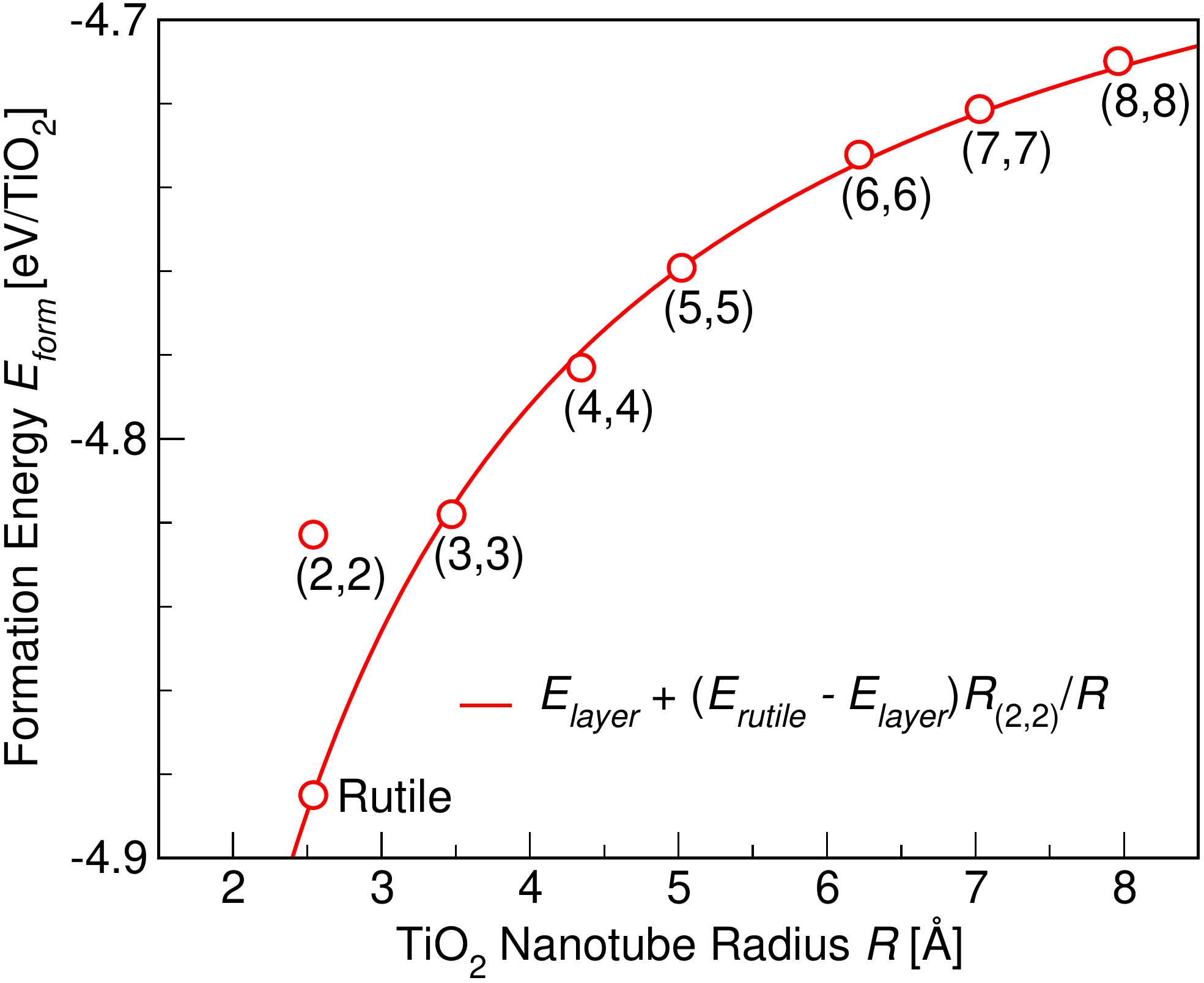}
\caption{
Formation energy $E_{f\!orm}$ in eV per TiO${}_{\text{2}}$ functional unit versus
radius $R$ in \AA~for TiO${}_{\text{2}}$ (2,2) nanorods and ($n,n$) nanotubes
($3\leq n \leq 8$) ({\color{red}{$\medcirc$}}, red). For comparison,
the TiO${}_{\text{2}}$ nanotube formation energies are also approximated by
$E_{f\!orm} \approx E_{layer} + (E_{rutile} - E_{layer})
R_{\text{(2,2)}}/R$, ({\color{red}{\textbf{---$\!$---$\!$---}}},
red), where $R_{\text{(2,2)}} \approx $2.54 \AA~is the radius of a
TiO${}_{\text{2}}$ (2,2) nanorod.}\label{TiO2_NT_Stability}

\end{figure}

From the formation energies, we see that (TiO${}_{\text{2}}$)${}_n$
clusters become generally more stable with increasing size, as
expected.  On the other hand, the formation energies of the ($n$,$n$)
TiO${}_{\text{2}}$ nanotubes ($n \leq$ 4) suggest that they are
all surprisingly stable, being more stable than the 2D structures and within
approximately 0.2 eV of the least stable bulk phases.  Even more
surprising, the TiO${}_{\text{2}}$ nanotubes seem to become
more stable with \emph{decreasing} size.   This is clearly seen in
Fig.~\ref{TiO2_NT_Stability}, which shows the radial dependence of the
formation energies for the TiO${}_{\text{2}}$ (2,2) nanorod and TiO${}_{\text{2}}$ ($n,n$)
nanotubes ($3\leq n \leq 8$).

In a simple classical picture, the strain energy to be overcome when
bending a layer into a tube should be inversely proportional to the
square of the tube radius ($\sim R^{-{2}}$). Such models have 
been shown to accurately describe TiO${}_{\text{2}}$ anatase and
lepidocrocite layer nanotubes \cite{TiO2LayerTBCalcs}.
However, the 
HexABC TiO${}_{\text{2}}$ nanotube system is complicated
by the fact that  
TiO${}_{\text{2}}$ is not found naturally in a HexABC layered structure.   
As a result, the strain energy introduced by bending the layer may be
overcome by changes in bonding which better resemble the bulk TiO${}_{\text{2}}$
structures.  This is seen in Fig.~\ref{Fig1}, where we find oxygen atoms have
$sp^{\text{3}}$ character in both the infinite layer structure and the
(4,4) TiO${}_{\text{2}}$ nanotube, but $sp^{\text{2}}$ character in the more
stable bulk rutile and (2,2) TiO${}_{\text{2}}$ nanorod.

As shown in Fig.~\ref{TiO2_NT_Stability}, the radial dependence of the
nanotube formation energies may be well approximated by 
\begin{eqnarray}
E_{f\!orm} &=& E_{layer} + (E_{rutile} - E_{layer}) R_{\text{(2,2)}} / R.\label{NTform}
\end{eqnarray}
Here \(E_{layer} \approx \) 4.63 eV/TiO${}_{\text{2}}$ is the formation energy
of the nanotube layer calculated using the optimized unit cell
parameters for the nanotubes, \(E_{rutile} \approx \) 4.89 eV/TiO${}_{\text{2}}$
is the formation energy of the bulk rutile phase, $R_{(\text{2,2})}$ is the
radius of a TiO${}_{\text{2}}$ (2,2) nanorod, and $R$ is the nanotube radius.
The first term in (\ref{NTform}) ensures the nanotube formation energy
tends to that for the infinite layer at large nanotube radii.  The
second term models the increase in stability as the nanotube is `bent'
into a more rutile-like structure at smaller radii. This reaches a
maximum for a (2,2) nanorod (\(R = R_{(\text{2,2})}\)), where the structure
becomes that of the rutile phase.

It should be noted that DFT formation energies have been shown to not
accurately describe the relative stability of anatase and rutile
phases, independent of xc-functional used \cite{Labat1}.  This should
be borne in mind when comparing formation energies. 

Comparison of the doped formation energies with those of the undoped
TiO${}_{\text{2}}$ structures shows that although less 
stable, doped forms of both TiO${}_{\text{2}}$ clusters and
nanotubes should be experimentally realizable.  This is
further justified by recent nitrogen doping experiments of large TiO${}_{\text{2}}$
nanotubes which yielded doping fractions between  1\% and
10\%~\cite{N-TiO2NTdoping,N-TiO2NTarraydoping,N-TiO2NTdoping2}.    

The formation energies of the doped systems also appear to correlate
well with both those of the undoped structures, and the relative
doping fraction.  Further, structures are typically less stable when
doping induces oxygen dislocations, which occur for boron and nitrogen
doped (TiO${}_{\text{2}}$)${}_{\text{7}}$ clusters, (TiO${}_{\text{2}}$)${}_{\text{8}}$ clusters, and bulk
phases. On the other hand, systems are more stable when the TiO${}_{\text{2}}$
structure is unchanged by the introduction of a dopant, as is the case
for nitrogen doping of (TiO${}_{\text{2}}$)${}_n$ clusters with surface sites
($n = 5,6,9$), nanotubes, layers, and surfaces.  Such a distinction
will also prove useful when we analyze the electronic structure of the
doped systems in Sec.~\ref{electronic}.

\section{Electronic Structure}\label{electronic}

\subsection{DOS of 0D Clusters}

\begin{figure}
\includegraphics[width=\columnwidth]{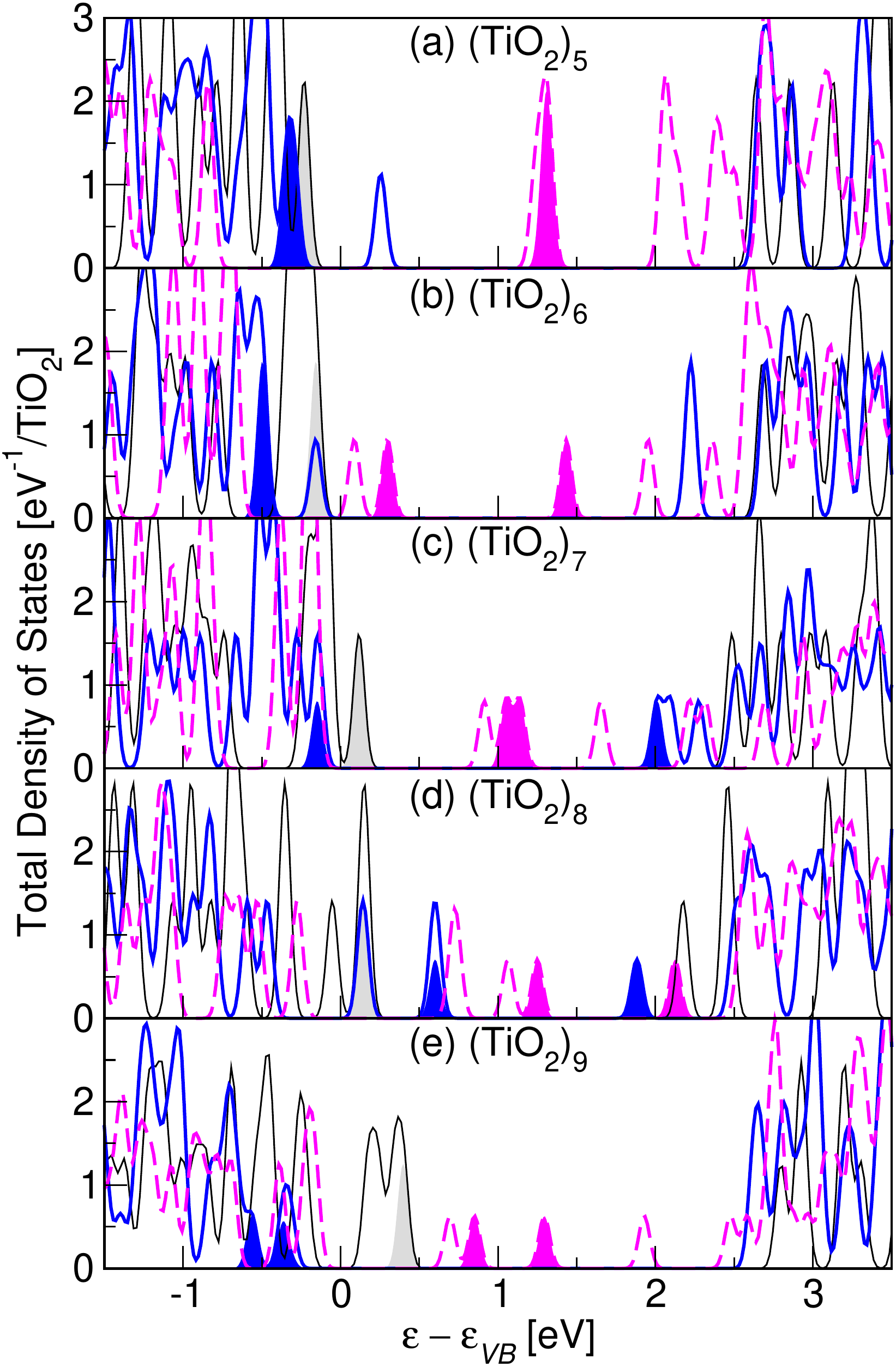}
\caption{Total density of states in eV$^{\text{-1}}$ per
  TiO${}_{\text{2}}$ functional unit versus energy in eV for undoped
  (thin black solid line), boron doped (magenta dashed line) and
  nitrogen doped (blue thick solid line) 
  (a--e) (TiO${}_{\text{2}}$)${}_n$ clusters for   5 $\leq n \leq$ 9, with
  doping fractions of   $\nicefrac{{1}}{\text{2}n}$. Energies
  are measured from the top of the valence bands of anatase TiO${}_{\text{2}}$
  $\varepsilon_{V\!B}$, and the DOSs are shifted to align the lowest
  eigenstate with that for anatase TiO${}_{\text{2}}$.  Occupancy is denoted
  by curve filling for states in the band gap.  
}\label{0D_DOS}
\end{figure}

We shall begin our analysis of the electronic structure of boron and
nitrogen doped TiO${}_{\text{2}}$ nanostructures by first considering
(TiO${}_{\text{2}}$)${}_n$ clusters (5$\leq n \leq $9).  As well as being of
interest in its own right, the study of such systems provides insight
into how local changes in nanostructure may change whether a dopant
behaves as a donor or acceptor site.  Such information will prove
useful when trying to form $p$-type TiO${}_{\text{2}}$ semiconducting
materials. 

\begin{figure}
\includegraphics[width=\columnwidth]{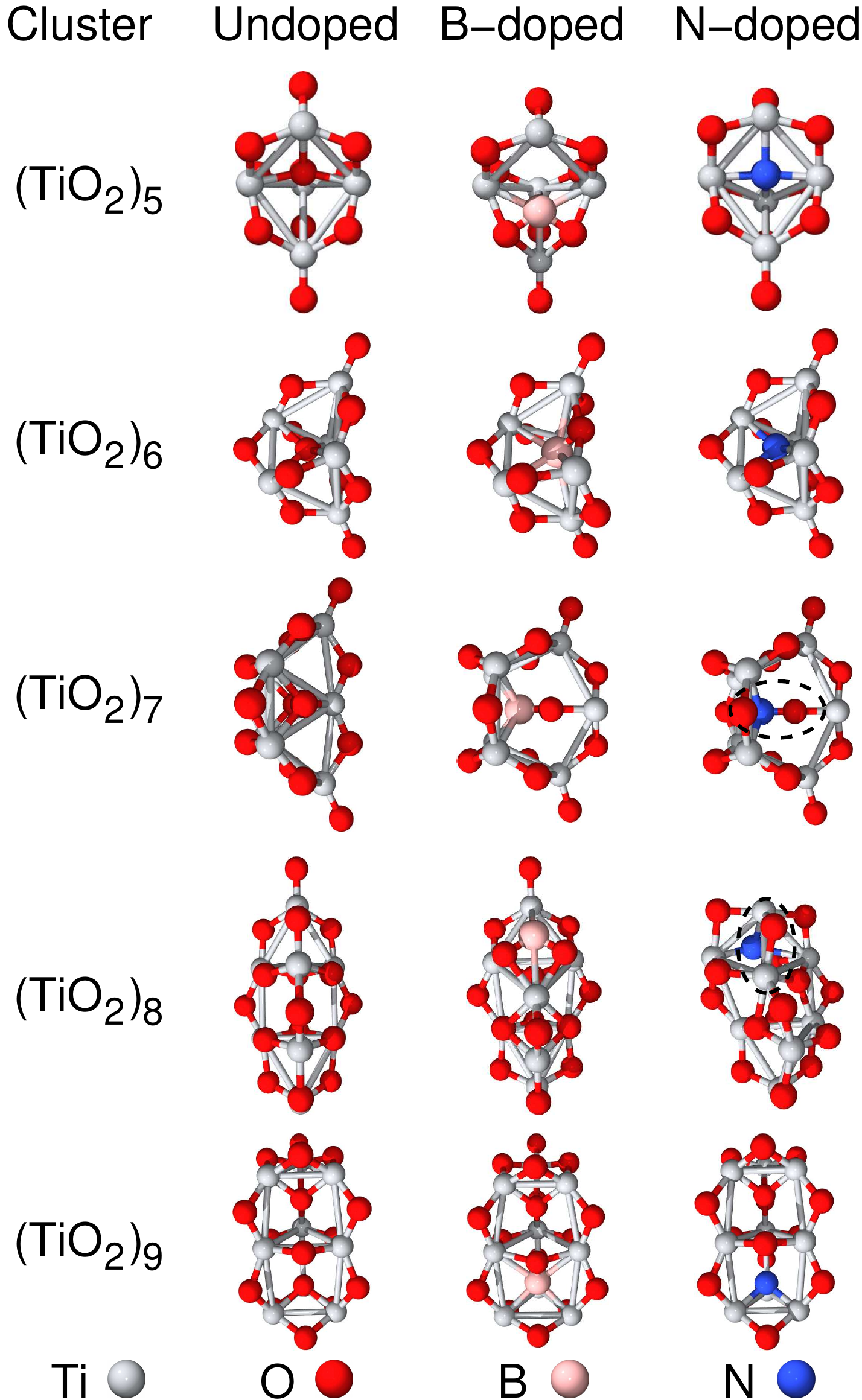}
\caption{Schematics of undoped (left), boron doped (middle), and nitrogen
    doped (right) (TiO${}_{\text{2}}$)${}_{n}$ clusters for 5 $\leq n \leq $ 9,
    with doping fractions of $\nicefrac{1}{2n}$. Note the circled
    nitrogen-oxygen bonds in the nitrogen doped structures for $n =$ 7
    and 8, where nitrogen acts as a donor~\cite{Structures}.  
}\label{0D_Structures}
\end{figure}

The DFT calculated DOS and structures for the most stable boron and
nitrogen doped (TiO${}_{\text{2}}$)${}_{n}$ clusters are shown in Figs.~\ref{0D_DOS}
and ~\ref{0D_Structures}, for 5 $\leq n \leq$ 9.  In general, boron
prefers to replace the most highly coordinated oxygen in TiO${}_{\text{2}}$
clusters, typically in a central location, and forms boron-oxygen
bonds via oxygen dislocations, as shown in Fig.~\ref{0D_Structures}.  Due
to its electropositive character, boron acts as a donor, introducing
three occupied  mid-gap states near the lowest unoccupied molecular
orbital (LUMO), as seen in Fig.~\ref{0D_DOS}.  

However, for (TiO${}_{\text{2}}$)${}_n$ clusters the influence of nitrogen
dopants is not as straightforward.  As with rutile TiO${}_{\text{2}}$
surfaces\cite{Nambu,Graciani}, 
nitrogen prefers to occupy sites which are 3-fold coordinated to
titanium.  For clusters consisting only of surface
sites ($n = $ 5, 6, 9), oxygen is tightly constrained, and no
significant changes in the cluster's structure occur via oxygen
dislocations, as shown in Fig.~\ref{0D_Structures}. As with rutile
TiO${}_{\text{2}}$ surfaces\cite{Nambu,Graciani}, nitrogen acts as an acceptor,
introducing an unoccupied mid-gap state near the highest occupied
molecular orbital (HOMO).    

For clusters where the nitrogen dopant's oxygen neighbors occupy
interior/bulk sites ($n =$ 7, 8), oxygen is more mobile, and may form
new nitrogen-oxygen bonds, as shown in Fig.~\ref{0D_Structures}. Since
oxygen  is more electronegative than nitrogen, nitrogen may transfer
charge to the bonding oxygen through these bonds. In this case,
nitrogen instead acts as a donor, introducing occupied states near the
LUMO. 

In general, we find when dopants form bonds with oxygen atoms via 
oxygen dislocations, they act as donor sites.  On the other hand, when
dopants such as nitrogen leave the TiO${}_{\text{2}}$ structure unchanged, they
act as acceptor sites, yielding a somewhat more stable structure.  We
shall find such a description useful in understanding why nitrogen
dopants behave differently in low dimensional systems and bulk
systems.

\subsection{DOS of 1D Nanorods and Nanotubes}

\begin{figure}
\includegraphics[width=\columnwidth]{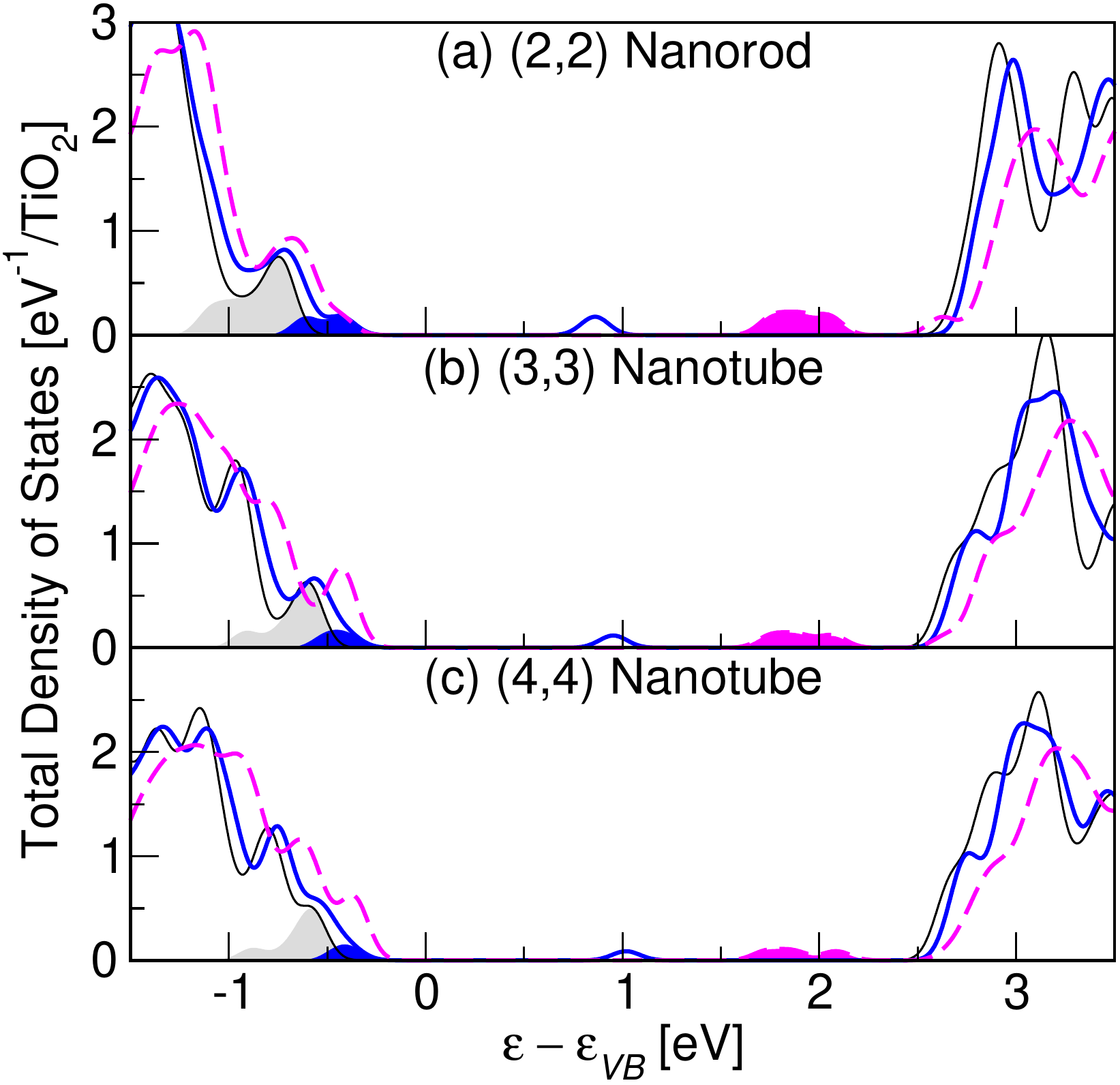}
\caption{RPBE calculation of the total density of states in eV$^{\text{-1}}$ per
  TiO${}_{\text{2}}$ functional unit versus energy in eV for undoped
  (thin black solid line), boron doped (magenta dashed line) and
  nitrogen doped (blue thick solid line) TiO${}_{\text{2}}$ (a) (2,2) nanorods,
  (b) (3,3) nanotubes and (c) (4,4) nanotubes, with 3.1\%,  2.1\%, and
  1.6\% doping respectively. Energies are measured from the top of the
  valence bands of anatase TiO${}_{\text{2}}$ $\varepsilon_{V\!B}$, and the
  DOSs are shifted to align the lowest eigenstate with that for
  anatase TiO${}_{\text{2}}$.  Occupancy is denoted by curve filling for states
  in the band gap.
}\label{1D_DOS}
\end{figure}

\begin{figure}
\includegraphics[width=\columnwidth]{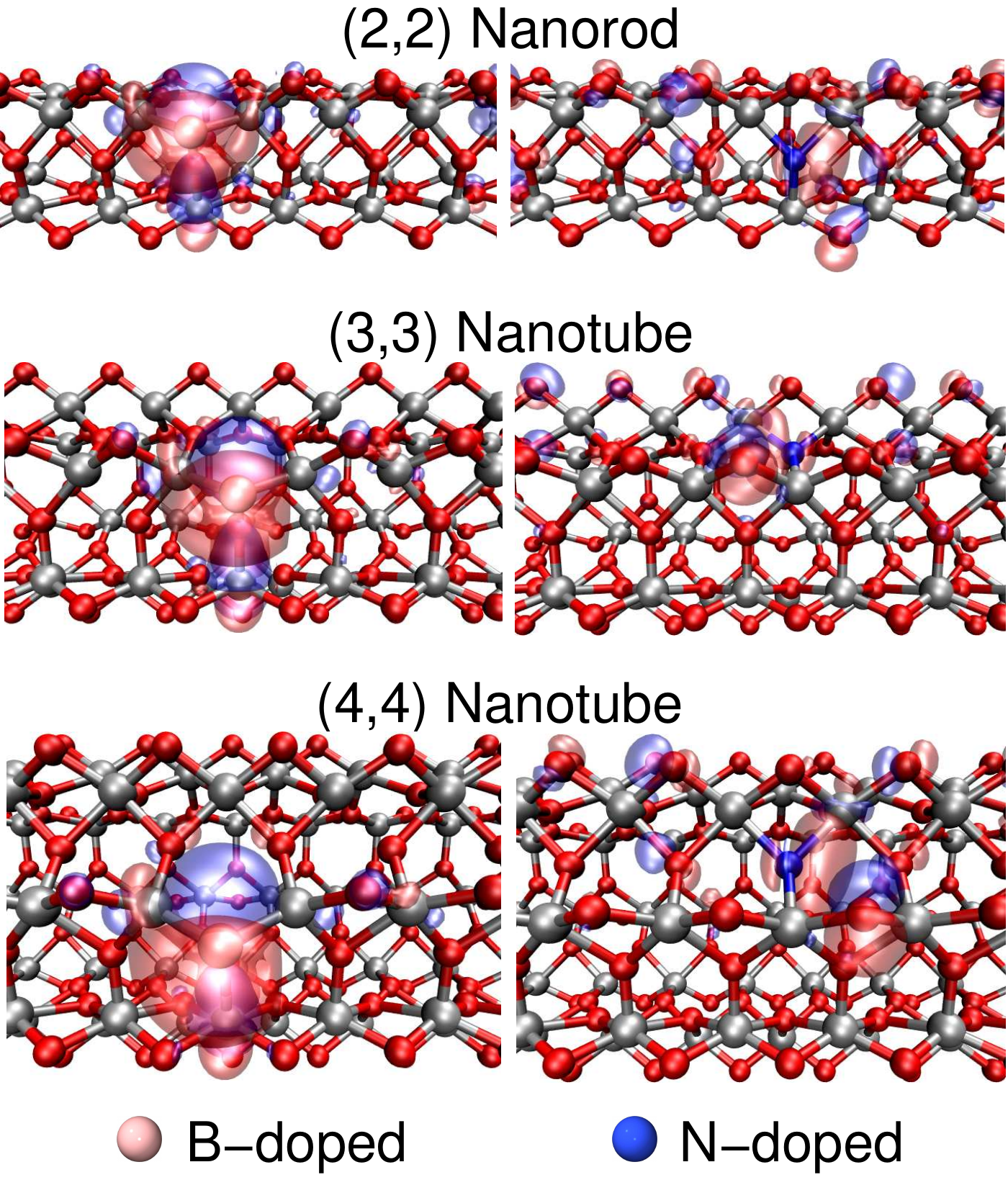}
\caption{Schematics of boron doped (left) and nitrogen doped (right)
  TiO${}_{\text{2}}$ (2,2) nanorods,  (3,3) nanotubes, and (4,4) nanotubes,
  with 3.1\%, 2.1\%, and 1.6\% doping respectively.  The highest
  occupied states for boron and nitrogen doped TiO${}_{\text{2}}$ 1D structures
  are depicted by isosurfaces of
  $\pm$0.05$e/$\AA$^{\text{3}}$.
}\label{1D_Structures} 
\end{figure}

\begin{figure}
\includegraphics[width=\columnwidth]{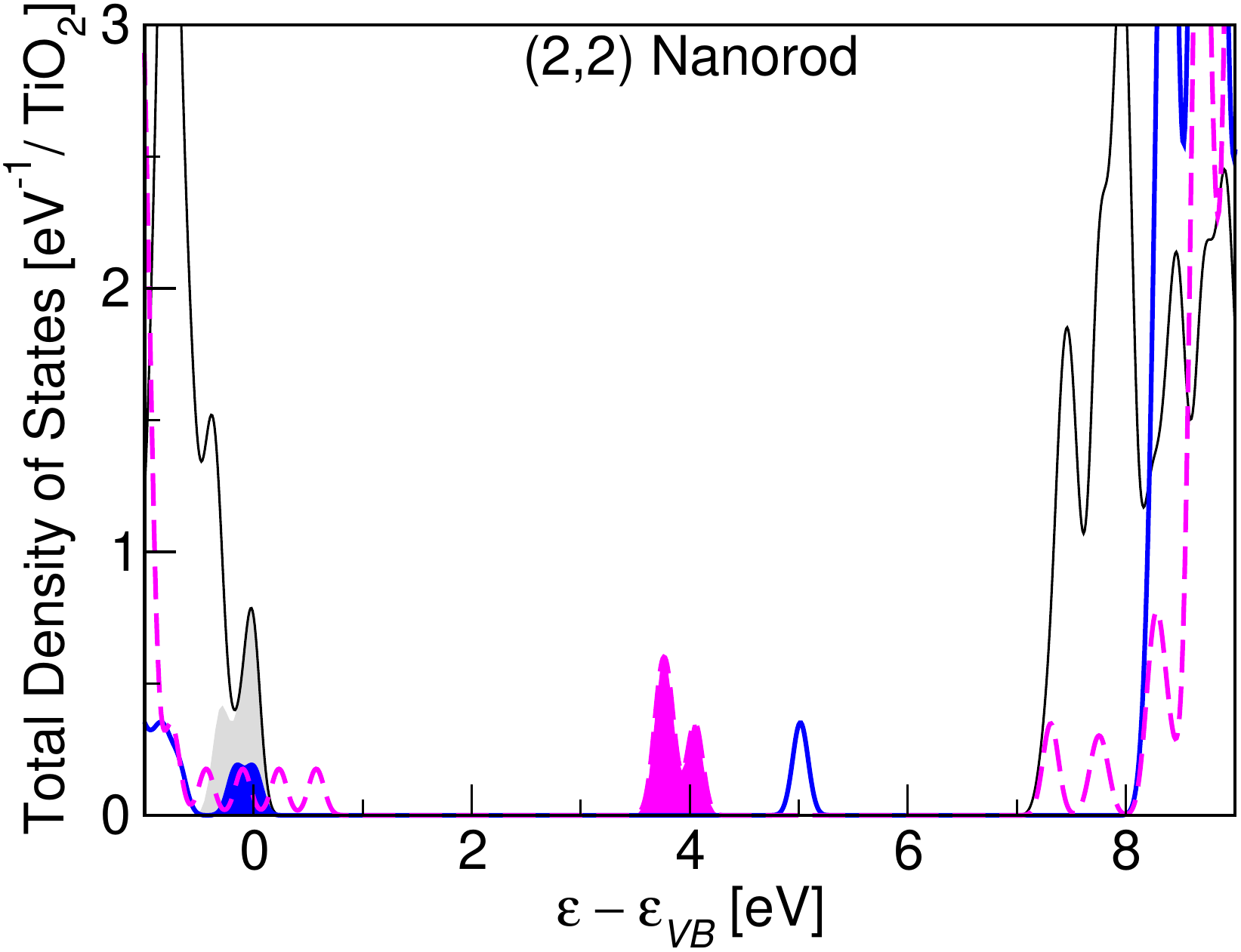}
\caption{GW calculation of the total density of states in eV$^{\text{-1}}$ per
  TiO${}_{\text{2}}$ functional unit versus energy in eV for undoped (thin
  black solid line), boron doped (magenta dashed line) and nitrogen
  doped (blue thick solid line) TiO${}_{\text{2}}$ (2,2) nanorods with 3.1\%
  doping. Energies are measured from the top of the  valence bands of
  the undoped TiO${}_{\text{2}}$ nanorod $\varepsilon_{V\!B}$, and the DOSs are
  shifted to align the lowest eigenstate with that for the undoped
  TiO${}_{\text{2}}$ nanorod.  Occupancy is denoted by curve filling for states
  in the band gap.
}\label{1D_DOS_GW}
\end{figure}

We shall now discuss how boron and nitrogen doping influences the
electronic structure of stable 1D nanorods and nanotubes.  Although
the the most stable doping sites are the same as those obtained for doped
TiO${}_{\text{2}}$ rutile surface and bulk \cite{Nambu,Graciani,TiO2PRL}, the influence of
dopants on the DOS is rather different. 

Figure \ref{1D_DOS} shows the DFT calculated DOS and Fig.~\ref{1D_Structures} the
structures of the most stable boron doped and nitrogen doped TiO${}_{\text{2}}$
(2,2) nanorods, (3,3) nanotubes, and (4,4) nanotubes.  The highest
occupied state is also shown as isosurfaces of
$\pm$0.05$e/$\AA${}^{\text{3}}$ in the side views of the doped structures.   

As with TiO${}_{\text{2}}$ clusters, the influence of boron dopants on
TiO${}_{\text{2}}$ nanorods and nanotubes may be understood in terms of boron's
weak electronegativity, especially when compared with the strongly
electronegative oxygen.  We find that boron prefers to occupy oxygen
sites which are 2-fold coordinated to neighboring titanium atoms.
However, as with the 0D clusters, boron's relatively electropositive 
character induces significant structural changes in the 1D structures,
creating a stronger third bond to a neighboring three-fold
coordinated oxygen via an oxygen dislocation, as shown in
Fig.~\ref{1D_DOS}. This yields three occupied mid-gap states localized on
the boron dopant, which overlap both the valence band O 2$p_\pi$ and
conduction band  Ti 2$d_{x y}$ states, as shown in Fig.~\ref{1D_DOS}. Boron
dopants thus yield donor states near the conduction band, which may be
photocatalytically active in the visible region.  However, the quantum
confinement inherent in these 1D structures may stretch these gaps, as
found for the GW calculated DOS shown in Fig.~\ref{1D_DOS_GW}. 

\begin{figure}
\includegraphics[width=\columnwidth]{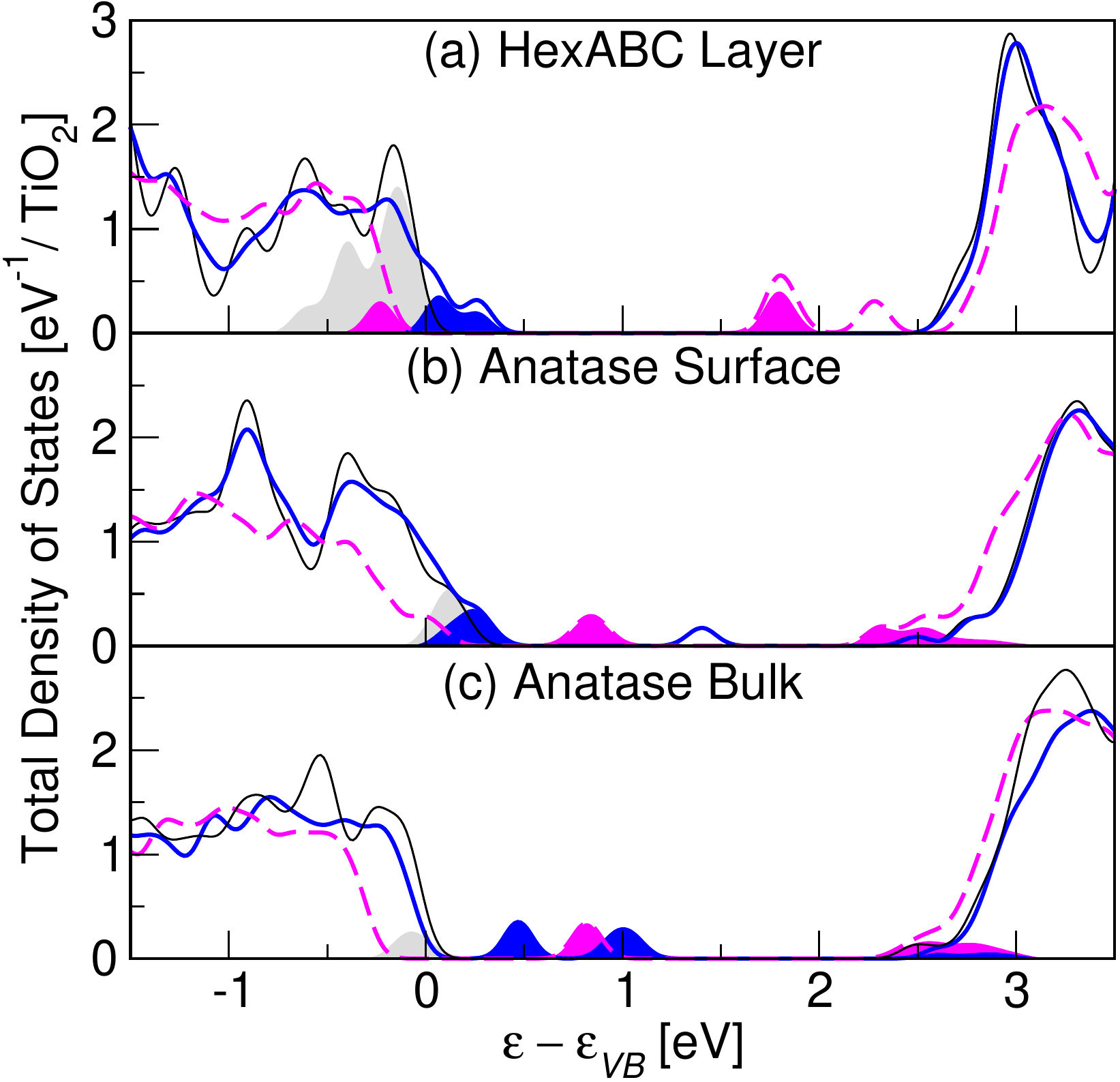}[!b]
\caption{Total density of states in eV$^{\text{-1}}$ per
  TiO${}_{\text{2}}$ functional unit versus energy in eV for undoped
  (thin black solid line), boron doped (magenta dashed line) and nitrogen doped (blue thick solid line)
  (a)
  TiO${}_{\text{2}}$ HexABC layer, (b) TiO${}_{\text{2}}$ anatase surface, and
  (c) bulk anatase, with 5.6\%, 3.1\%, and 3.1\% doping
  respectively. Energies are measured from the top of the 
  valence bands of anatase TiO${}_{\text{2}}$ $\varepsilon_{V\!B}$, and the DOSs are shifted
  to align the lowest eigenstate with that for anatase
  TiO${}_{\text{2}}$.  Occupancy is denoted by curve filling for states
  in the band gap. 
}\label{2D3D_DOS}
\end{figure}

On the other hand, we find nitrogen dopants prefer to occupy oxygen sites
which are 3-fold coordinated to Ti, as was previously found for the
rutile TiO${}_{\text{2}}$ surface\cite{Nambu, Graciani}.  This yields one
occupied state at the top of the valence band and one unoccupied
mid-gap state in the same spin channel. Both states are localized on
the nitrogen dopant but overlap the valence band O 2$p_\pi$ states, as
shown in Fig.~\ref{1D_Structures}.  Nitrogen dopants thus act as acceptors,
providing localized states well above the valence band, as is also
found for the GW calculated DOS shown in Fig.~\ref{1D_DOS_GW}. 

Although we find nitrogen dopants act as acceptors in TiO${}_{\text{2}}$ 1D
structures, such large gaps between the valence band and the
unoccupied mid-gap states would not yield $p$-type semiconductors.
This may be attributed to the substantial quantum confinement in these
1D structures.  However, for 2D and 3D systems, we find it is possible
to produce both $p$-type and $n$-type classical semiconductors.

\subsection{DOS of 2D Layers,  3D Surfaces and Bulks}

\begin{figure}
\includegraphics[width=\columnwidth]{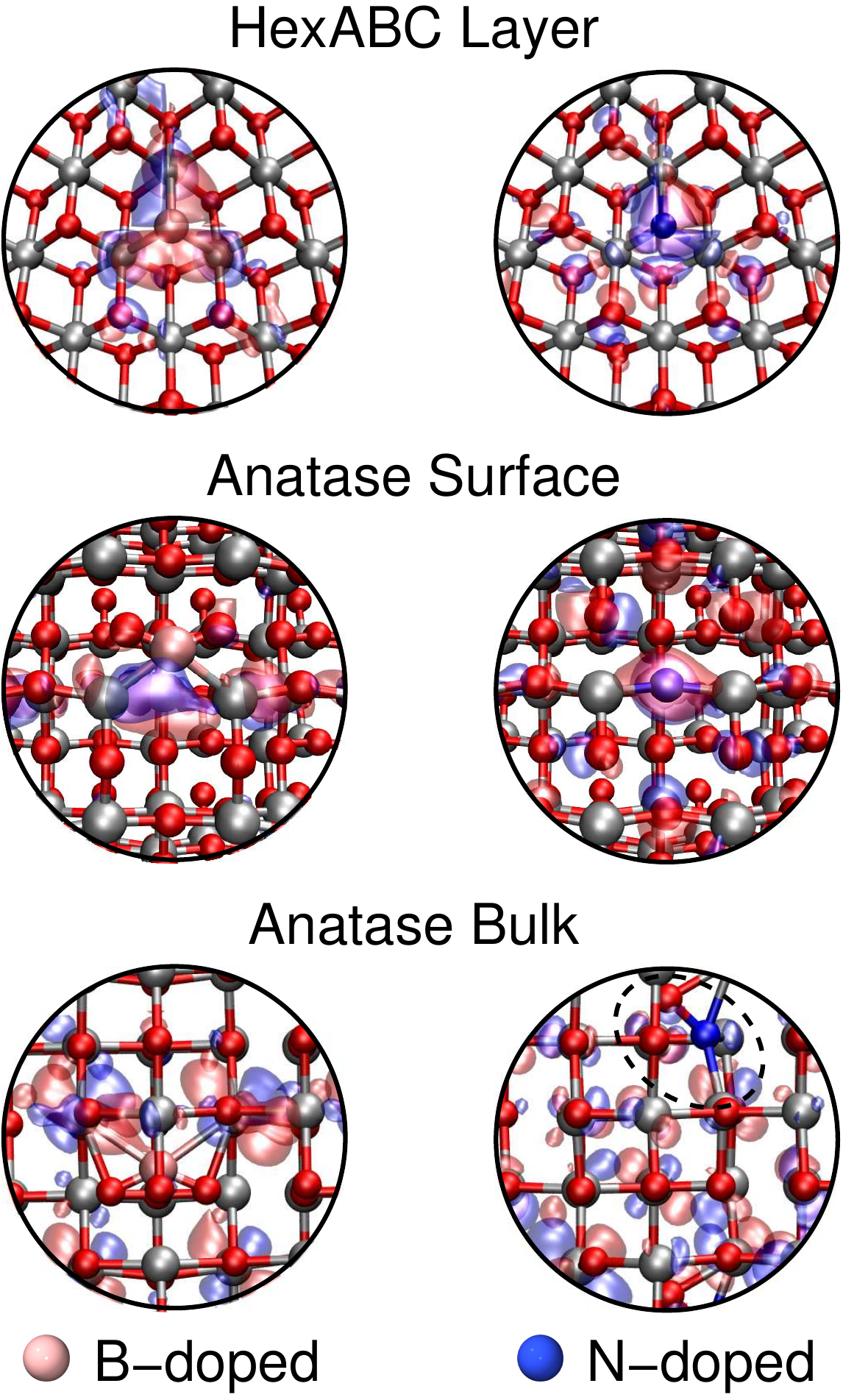}
\caption{Schematics of the boron doped (left) and nitrogen doped (right) 
 HexABC layer, anatase surface, and  anatase bulk
   phases, with 5.6\%, 3.1\%, and 3.1\% doping respectively.
 The highest occupied states for boron and nitrogen doped TiO${}_{\text{2}}$ 2D and 3D
 structures are depicted by isosurfaces of
 $\pm$0.05$e/$\AA$^{\text{3}}$. Note the circled nitrogen-oxygen bonds in the nitrogen doped
 structures for the bulk phase, where nitrogen acts as a
 donor~\cite{Structures}. 
}\label{2D3D_Structures}
\end{figure}

To determine the reliability of our DFT and GW electronic structure
calculations, we now consider the influence of boron and nitrogen doping on the
DOS for 2D and 3D TiO${}_{\text{2}}$ structures, which may be compared with
recent experiments.

For the HexABC layer shown in Figs.~\ref{2D3D_DOS}(a) and
  \ref{2D3D_Structures}, boron and nitrogen 
dopants act as donors and acceptors respectively, as was the case for
the TiO${}_{\text{2}}$ nanotubes. However, for these 2D
structures the doping states are 
  sufficiently close in energy to the conduction band or valence band to allow 
  charge transfer for classical $n$-type or $p$-type doping by boron or nitrogen
  dopants respectively.  A similar behavior is found for 
the anatase surface, shown in Fig.~\ref{2D3D_DOS}(b) and
    Fig.~\ref{2D3D_Structures}. 
This suggests boron and nitrogen doped 2D
  TiO${}_{\text{2}}$ layers or surfaces may potentially be used for
  TiO${}_{\text{2}}$ based  electronics.  

On the other hand, for bulk anatase boron and nitrogen dopants both yield $n$-type
semiconductors, as is seen experimentally \cite{N-TiO2doping}.  We
attribute this difference to the nitrogen-oxygen bond shown in
Fig.~\ref{2D3D_DOS}(c), which may arise due to oxygen's greater structural
flexibility in the bulk. As a result, boron and nitrogen doping both yield mid-gap
  occupied Ti 2$d_{x y}$ conduction band states with little weight on the doping sites. Here,
  the dopants donate their valence electrons fully to the conduction band, as
  seen in Fig.~\ref{2D3D_DOS}(c). 

Nitrogen thus acts as a donor for bulk-like clusters and anatase,
in agreement with recent  findings for nitrogen doping of both bulk anatase
and thick bulk-like anatase nanopores~\cite{N-TiO2doping}.  This
suggests that $n$-type or $p$-type semiconductors may be
produced by nitrogen doping depending on whether nearby oxygen atoms occupy
lower-coordinated surface or higher-coordinated bulk sites.

Further, we also find that the highest occupied states for boron and nitrogen doped
systems become increasingly localized as the structure is
dimensionally constrained from
bulk $\rightarrow$ surface $\rightarrow$ layer $\rightarrow$ nanotube
$\rightarrow$ nanorod.
This may have important consequences for the photocatalytic activity
of these more localized states in lower dimensional structures.
On the other hand, this may be partially alleviated by having these
states located on  the structure's surface.  In this way,
electron-hole pairs need not travel significantly, as is the case for
bulk materials.

\subsection{Energy Gaps}

To analyze the qualitative trends in photocatalytic activity of boron and
nitrogen doped TiO${}_{\text{2}}$ nanostructures, we shall now compare the energy gaps
obtained using different approximations within DFT, GW calculations,
and experiment.  

\begin{figure}
\includegraphics[width=\columnwidth]{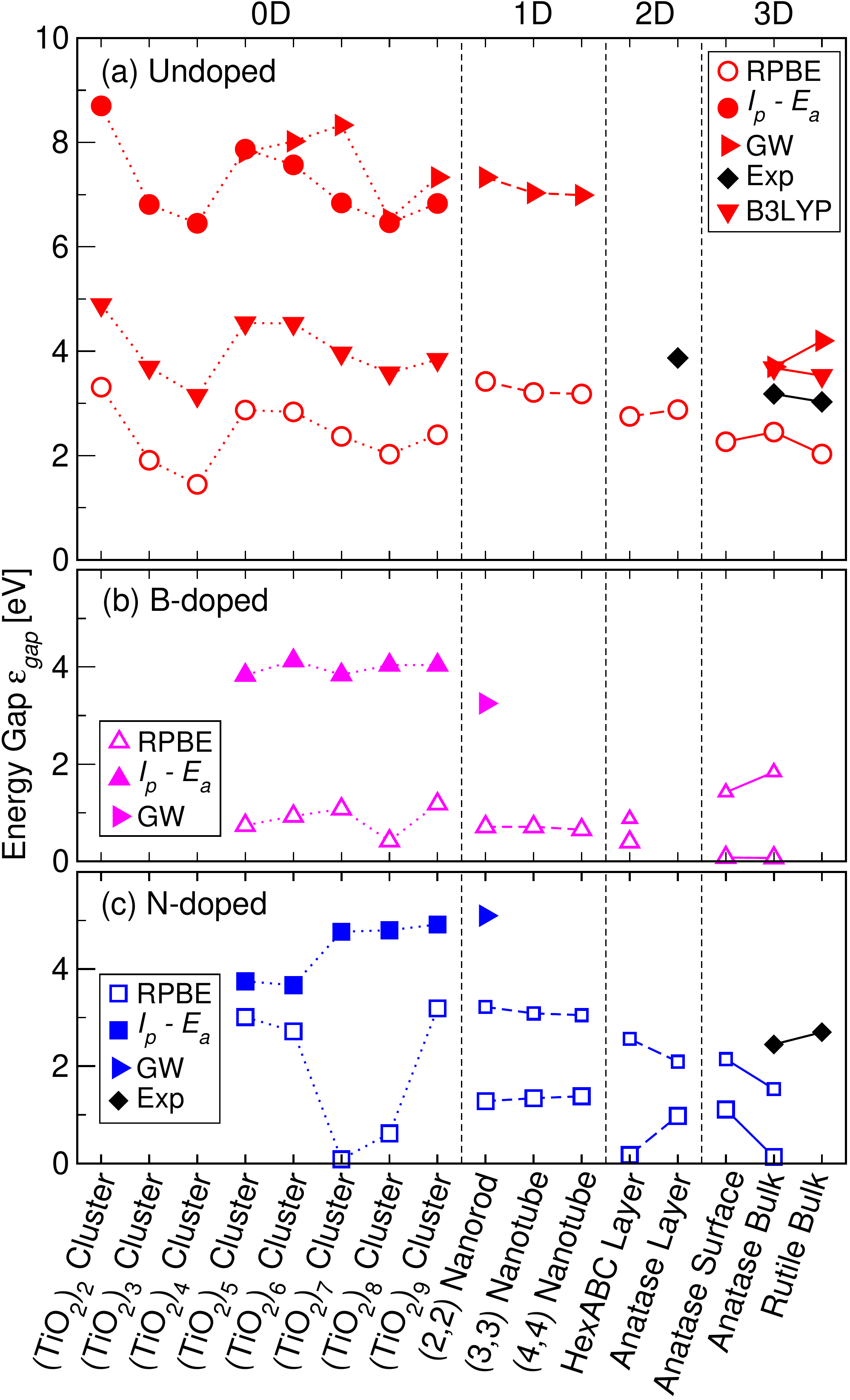}
\caption{Energy gap $\varepsilon_{gap}$  in eV versus TiO${}_{\text{2}}$
  structure for 0D (TiO${}_{\text{2}}$)${}_n$ clusters ($n\leq \text{9}$), 1D
  TiO${}_{\text{2}}$ (2,2) nanorods, (3,3) nanotubes, (4,4) nanotubes, 2D
  HexABC and anatase layers, and 3D anatase surface, anatase bulk, and
  rutile bulk phases. DFT calculations using RPBE of the highest
  occupied and lowest unoccupied state gaps / $I_p - E_a$ for the (a)
  undoped ({\color{red}{$\medcirc$}}/{\color{red}{\medbullet}}, red), (b)
  boron doped
  ({\color{magenta}{$\vartriangle$}}/{\color{magenta}{$\blacktriangle$}}, 
  magenta), and (c) nitrogen doped
  ({\color{blue}{$\boxempty$}}/{\color{blue}{$\blacksquare$}}, blue)
  systems are compared with undoped B3LYP
  ({\color{red}{$\blacktriangledown$}}), GW ($\blacktriangleright$),
  and experimental ($\Diamondblack$)
  results~\cite{Structures,GW_TiO2_anatase,N-TiO2NTarraydoping,TiO2Rev1}. 
  Small open symbols denote transitions between highest fully occupied
  states and the conduction band.         
}\label{Fig3} 
\end{figure}

The DFT calculated energy gaps $\varepsilon_{gap}$,  for
(TiO${}_{\text{2}}$)${}_{n}$ clusters ($n \leq $ 9) and band gaps for TiO${}_{\text{2}}$
(2,2) nanorods, (3,3) nanotubes, and (4,4) nanotubes, in the undoped,
boron doped, and nitrogen doped forms, are shown in Fig.~\ref{Fig3}.  For
the clusters, the gap is estimated by both the difference in energy
between the HOMO and LUMO, and the difference between the ionization
potential and electron affinity  energies $I_p - E_a$.  The band gap
for TiO${}_{\text{2}}$ nanorods, nanotubes, layers, surfaces and bulk phases is
approximated by the indirect band gap between highest occupied and
lowest unoccupied states. Although not as relevant for photoabsorption
as the direct gap, the large size required for the doped super cells,
shown in Fig.~\ref{1D_Structures} and Fig.~\ref{2D3D_Structures}, leaves
indirect and direct gaps indistinguishable. For the undoped nanorods
and nanotubes and bulk rutile we find the band gap is direct, while
for the undoped HexABC layer and anatase structures we find the band
gap is indirect.  For the same structures we also provide B3LYP band
gaps, GW calculations, and experimental results for comparison.    

As previous studies have shown, standard DFT tends to
underestimate band gaps for bulk TiO${}_{\text{2}}$ by approximately 1 eV,
due in part to self-interaction errors \cite{BandGapErrors,NewXCFuncs}. 
This may be partially addressed by the use of hybrid functionals such
as B3LYP, which generally seem to improve band gaps for bulk systems
\cite{Muscat, DeAngelis,DiValentin}. However, B3LYP calculations for
TiO${}_{\text{2}}$ clusters largely underestimate the gap relative to the more
reliable $I_p - E_a$, while for the bulks the gap is overestimated by
approximately 0.4 eV relative to experiment. Also, B3LYP
and RPBE calculations provide the same qualitative description of the
\emph{trends} in the energy gaps for TiO${}_{\text{2}}$.     

GW is probably the most successful and generally applicable method to
calculate quasi-particle gaps.  For clusters it agrees well with $I_p
- E_a$, but for bulk systems GW seems to overestimate the experimental
gap like B3LYP.  This overestimation by GW may be attributed to
excitonic effects, which are not included in GW, or the role played by
oxygen defects \cite{Hardman1994}.  For these reasons, GW and DFT
calculations are best used as upper and lower bounds for optical band
gaps.   

Fig.~\ref{Fig3} shows that for both 3D and 2D
systems, RPBE gaps underestimate the experimental results by
approximately 1 eV.  For 1D and 0D systems, we
find a much larger difference of about 4 eV and 5 eV respectively,
between the RPBE gaps and the $I_p-E_a$ and GW results.  We may
attribute this increasing disparity to the greater quantum confinement
and charge localization in the 1D and 0D systems, which yield greater
self-interaction effects.  We find B3LYP gaps also underestimate this
effect, simply increasing the RPBE energy gaps for both 0D and 3D
systems up by about 1.4 eV.     

On the other hand, we find the RPBE gaps reproduce qualitatively
the structural dependence of the $I_p - E_a$, GW, and experimental
results for a given dimensionality, up to a constant shift.  This is
true even for 3D bulk systems, where standard DFT does not predict
rutile to be the most stable \cite{Labat1}, as shown in Fig.~\ref{Fig1}. 

Whether calculated using RPBE, $I_p-E_a$ or GW, the energy gaps for
both boron and nitrogen doped TiO$_{\text{2}}$ nanostructures are generally
narrowed, as shown in Fig.~\ref{Fig3}(b) and (c).  However, for nitrogen
doped (TiO${}_{\text{2}}$)$_n$ clusters where nitrogen acts as an acceptor ($n
= 5, 6, 9$), the energy gap is actually \emph{increased} when spin is
conserved, compared to the undoped clusters in RPBE.  This effect is
not properly described by the $N\rightarrow N+1$ transitions of
$I_p-E_a$, for which spin is not conserved for these nitrogen doped
clusters.  On the other hand, when nitrogen acts as a donor ($n=7,8$)
the smallest gap between energy levels does conserve spin.    

The boron doped TiO$_{\text{2}}$ nanorods and nanotubes have perhaps the most
promising energy gap results of the TiO$_{\text{2}}$ structures considered
herein, as seen in Fig.~\ref{1D_DOS}(b).  Boron dopants introduce in the
nanorods localized occupied states near the conduction band edge in
both RPBE (\emph{cf.} Figs.~\ref{1D_DOS} and \ref{1D_Structures}) and GW
(\emph{cf.} Fig.~\ref{1D_DOS_GW})  calculations.  On the other hand,
nitrogen doping of nanorods and nanotubes introduces well defined
mid-gap states, as shown in Fig.~\ref{1D_DOS}(c).  However, to perform
water dissociation, the energy of the excited electron must be above
that for hydrogen evolution, with respect to the vacuum level.  This
is not the case for such a mid-gap state.  This opens the possibility
of a second excitation from the mid-gap state to the conduction band.
However, the cross section for such an excitation may be rather low. 

For boron doping of 2D and 3D structures, the highest occupied state
donates its electron almost entirely to the conduction band, yielding
an $n$-type semiconductor.  Thus at very low temperatures, the RPBE
band gap is very small.  The same is true for $n$-type nitrogen doped
bulk anatase.  For these reasons, we also provide the energy gap
between the highest fully occupied state and the conduction band,
which may be more relevant for photoabsorption.  We find these RPBE
gaps are still generally smaller than those for their undoped TiO$_{\text{2}}$
counterparts. 

In summary, for both boron and nitrogen doped clusters we find RPBE
gaps differ from $I_p - E_a$  by about 3 eV, while for nitrogen doped
anatase the RPBE gap differs from experiment by about 0.6 eV.  Given
the common shift of 1 eV for undoped 2D and 3D structures, this
suggests that both boron and nitrogen doped 2D TiO${}_{\text{2}}$ structures
are promising candidates for photocatalysis.  Further, the boron and
nitrogen doped 1D nanotube results also  warrant further experimental
investigation.

\section{Conclusion}

In conclusion, we have demonstrated how the electronic properties of 
TiO${}_{\text{2}}$ may be ``tailored'' using nanostructural changes in
combination with boron and nitrogen doping.  While boron doping tends
to produce smaller band gap $n$-type semiconductors, nitrogen doping produces
$p$-type or $n$-type semiconductors depending on whether or not
nearby oxygen atoms occupy surface sites.  This suggests that a $p$-type
TiO${}_{\text{2}}$ semiconductor may be produced using nitrogen doping in conjunction
with surface confinement at the nanoscale.  

\acknowledgments

We thank S.~In, Z.-W.~Qu, G.-J.~Kroes,  T.~Jaramillo, and
J.~K.~N{\o}rskov for useful discussions.  The authors acknowledge
financial support from NABIIT and the Danish Center for Scientific
Computing. J.I.M.~acknowledges the financial support of the STREP EU 
APOLLON-B Project through grant No.~NMP3-CT-2006-033228.  The Center
for Atomic-scale Materials Design (CAMD) is sponsored by the Lundbeck
Foundation.


\end{document}